\documentclass[10pt,letterpaper]{article}
\usepackage{opex3}

\usepackage{graphicx}
\graphicspath{{pictTriangular/}{pict/}{}}

\begin{document}

\emph{\textbf{\small 
Paper published in Optics Express, Vol. 15, Issue 19,
pp. 12380-12385 (2007).}}

%%%%%%%%%%%%%%%%%% title page information %%%%%%%%%%%%%%%%%%
\title{Low-threshold bistability
of slow light in photonic-crystal waveguides}

\author{S.~F.~Mingaleev,$^{1}$ A.~E.~Miroshnichenko,$^{2}$
and Yu.~S.~Kivshar$^{2}$}

\address{$^{1}$ Bogolyubov Institute for Theoretical Physics of the National
Academy of Sciences of Ukraine, 14-B Metrologichna Street, Kiev
03680, Ukraine\\

$^{2}$ Nonlinear Physics Centre and Centre for Ultra-high bandwidth
Devices for Optical Systems (CUDOS), Research School of Physical
Sciences and Engineering,\\
Australian National University, Canberra ACT 0200, Australia}

\email{mingaleev@bitp.kiev.ua, aem124@rsphysse.anu.edu.au}

\begin{abstract}
We analyze the resonant transmission of light through a
photonic-crystal waveguide side coupled to a Kerr nonlinear cavity,
and demonstrate how to design the structure geometry for achieving
bistability and all-optical switching at \emph{ultralow} powers in
the \emph{slow-light} regime. We show that the resonance quality
factor in such structures scales inversely proportional to the group
velocity of light at the resonant frequency and thus grows
indefinitely in the slow-light regime. Accordingly, the power
threshold required for all-optical switching in such structures
scales as a square of the group velocity, rapidly vanishing in the
slow-light regime.
\end{abstract}


%%%%%%%%%%%%%%%%%%%%%%% References %%%%%%%%%%%%%%%%%%%%%%%%%%%%%%%%%%%

%%%%%%%%%%%%%%%%%%%%%%%%%%%%%%%%%%%%%%%%%%%%%%%%%%%%%%%%%%%%%%%%%%%%%

\section{Introduction}

Recent studies of slow light and its properties are motivated by the
enhancement of light-matter interactions for smaller group
velocities leading to the enhancement of optical
gain~\cite{Dowling:1994-1896:JAP} and electro-optic
effect~\cite{Roussey:2006-241110:APL}, a growth of the spontaneous
emission rate~\cite{Suzuki:1995-570:JOSB}, and more efficient
nonlinear optical response
\cite{Scalora:1994-1368:PRL,Martorell:1997-702:APL,Soljacic:2002-2052:JOSB,Chen:2004-3353:OE}.
The concept of slow light is also useful for realizing all-optical
routers and optical buffers for pulse storage and synchronization.

Different concepts and schemes for realizing the slow-light
propagation in various media and structures have been suggested so
far. Although the most dramatic reduction of the group velocity of
light has been achieved in atomic media and it is based on
electromagnetically-induced transparency, such media are not
suitable for high-bit-rate optical systems due to their high
dispersion~\cite{Khurgin:2005-1062:JOSB}. In contrast, alternative
realizations of the slow-light propagation in high-index-contrast
coupled resonator structures can be very useful for creating optical
buffers operating at hundreds of
gigabits/s~\cite{Khurgin:2005-1062:JOSB}. Recently, ultra-compact
coupled-resonator optical buffers on a silicon chip have been
experimentally fabricated with a large fractional group delay
exceeding 10 bits, achieved for bit rates as high as 20
gigabits/s~\cite{Xia:2007-65:NATPHOT}.

\begin{figure}[t]
\centering\includegraphics[width=11cm]{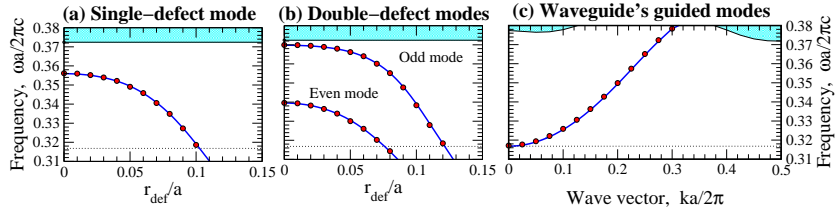}
\caption{Frequencies of localized cavity modes created by changing
the radius $r_{\rm def}$ of (a) a single defect rod, and (b) two
neighboring defect rods in the photonic crystal created by a
triangular lattice of rods with $\varepsilon=12$ and radius
$r=0.25a$ in air, $a$ is the lattice spacing. (c) Dispersion of the
W1 photonic-crystal waveguide created by removing a row of rods in
the same photonic crystal. Results are calculated using eleven
maximally localized Wannier functions \cite{Busch:2003-R1233:JPCM}
(blue lines) in an excellent agreement with the supercell
plane-waves method \cite{mpb} (red circles).} \label{fig:fig1}
\end{figure}

Compact coupled-resonator optical systems can be realized on the
basis of photonic-crystal waveguides, for which the slow-light
propagation with the smallest achieved group velocity reaching
$c/1000$ has been experimentally demonstrated
\cite{Notomi:2001-253902:PRL,Jacobsen:2005-7861:OE,Vlasov:2005-65:NAT,Gersen:2005-073903:PRL}.
Because of this success, the interest to the slow-light applications
based on photonic-crystal waveguides is rapidly growing, attracting
attention to the problems of designing waveguide
bends~\cite{Assefa:2006-745:OL}, couplers~\cite{Vlasov:2006-50:OL},
and other types of functional optical devices which would
efficiently operate in the slow-light regime.

Dynamic control of the slow-light propagation can be realized by
direct tuning of the group velocity, either
thermo-electrically~\cite{Vlasov:2005-65:NAT} or
all-optically~\cite{Mok:2006-775:NATPHYS}. The scales of the
corresponding optical devices are, however, sub-optimal.
Potentially, they can be reduced dramatically in the
waveguide-cavity structures with tunable high-quality
resonances~\cite{Fan:2002-908:APL}. Using a nonlinear cavity, active
switching and other functionalities of such devices can be realized
by shifting the resonance frequency all-optically, e.g. by changing
the power of the incoming light in order to achieve the bistable
transmission. Several successful experimental realizations of
low-threshold light switching in such structures have been recently
reported~\cite{Almeida:2004-1081:NAT,Barclay:2005-801:OE,Notomi:2005-2678:OE,
Tanabe:2005-151112:APL,Priem:2005-9623:OE,Uesugi:2006-377:OE,Yang:2007-0703132:arXiv}.
However, none of those demonstrated devices could operate in the
slow-light regime.

Making the waveguide-cavity structures to be applicable and useful
for the dynamic control of the slow-light propagation is a
nontrivial task, due to strong extrinsic scattering losses caused by
most types of side-coupled cavities (including those introduced by
fabrication imperfections) for operating frequencies close to the
edges of the propagation
bands~\cite{Hughes:2005-033903:PRL,Mingaleev:2006-046603:PRE}.

In this paper, we demonstrate how to employ the recently suggested
geometry-based enhancement of the resonance quality
factor~\cite{Mingaleev:2006-046603:PRE} to design efficient
waveguide-cavity structures for ultralow-threshold bistability and
all-optical switching in the slow-light regime.

The outline of the paper is as follows. Section 2 describes the
model we adopt, using the single-defect geometry shown in
Fig.~\ref{fig:fig2} as a concrete example, and demonstrates why such
type of geometry exhibits the resonance quality factor $Q \sim v_g$
which scales linearly with the group velocity of light, $v_{g}$, and
thus vanishes at \emph{both} propagation band edges. Section 3
describes the contrasting mechanism of slow light scattering which
leads to $Q \sim 1/v_{g}$ growing indefinitely at one of the
propagation band edges. Correspondingly, the power threshold $P_{\rm
th}$ required for all-optical light switching decreases as $P_{\rm
th} \sim Q^{-2} \sim v_{g}^2$ in this case. We illustrate this
mechanism on the example of the double-defect geometry shown in
Fig.~\ref{fig:fig3}. Section 4 concludes the paper.

\section{Model and the basic parameters}

\begin{figure}[t]
\centering\includegraphics[width=12cm]{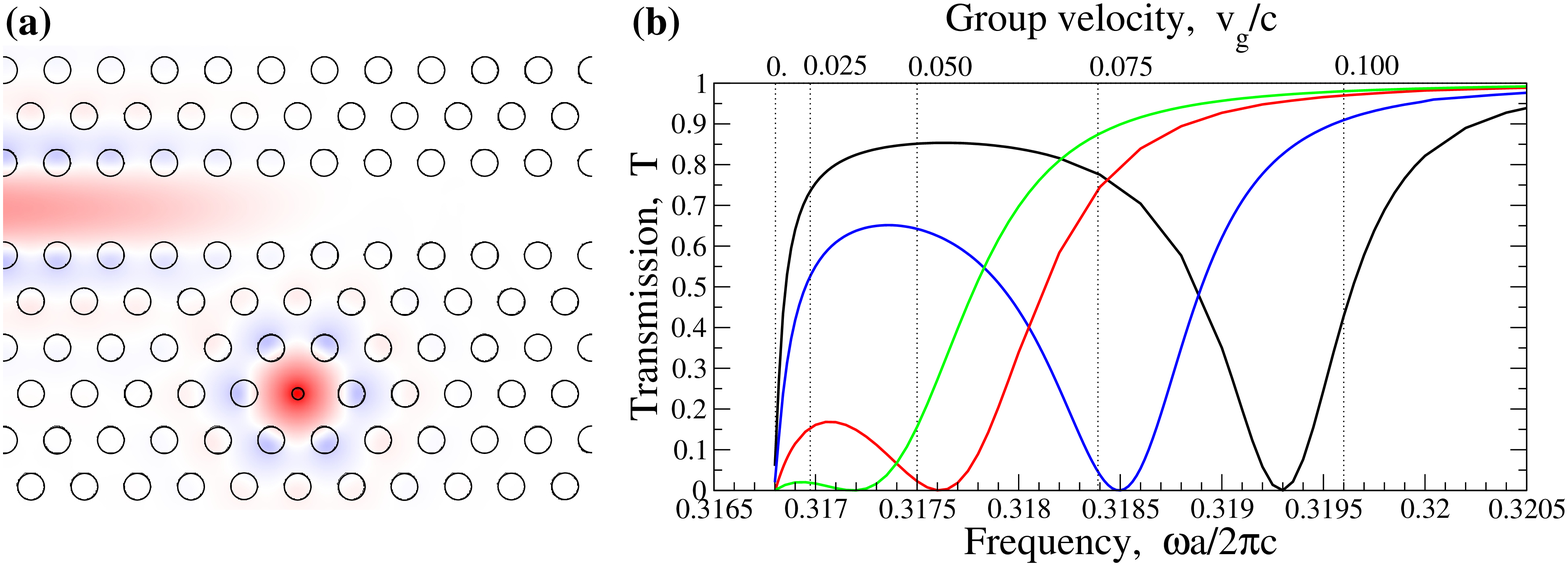}
\caption{Single-defect waveguide-cavity structure with the radius of
the defect rod $r_{\rm{def}}$: (a) Electric field at the resonance
reflection for $r_{\rm{def}}=0.102a$; (b) Transmission spectra for
different values of $r_{\rm{def}}$: $0.1a$ (black), $0.101a$ (blue),
$0.102a$ (red), $0.1025a$ (green). For convenience, in addition to
the light frequency on the bottom axis, we indicate on the top axis
the complementary group velocity, $v_{g}(\omega)$, of the
waveguide's guided mode.} \label{fig:fig2}
\end{figure}

To illustrate our basic idea, we consider the simplest case of a
two-dimensional photonic crystal (PhC) created by a triangular
lattice of dielectric rods in air. The rods are made of either Si or
GaAs ($\varepsilon=12$) with the radius $r=0.25a$,  $a$ is the
lattice spacing. Such photonic crystal has two large band gaps for
the E-polarized light (electric field parallel to the rods), and we
use the first gap between the frequencies $\omega a/2 \pi c =
0.2440$ and $0.3705$. By reducing the radius of {\em a single rod},
we create a monopole-like localized defect mode in this bandgap [see
Fig.~\ref{fig:fig1}(a)]. Reducing the radius of {\em two neighboring
rods} allows to create two localized modes, one with odd and another
with even field symmetry [see Fig.~\ref{fig:fig1}(b)]. Removing a
row of rods creates the so-called W1-waveguide which guides light
with the frequencies $\omega(k)$ determined by the guided mode wave
vector $k$ as shown in Fig.~\ref{fig:fig1}(c). The group velocity
$v_{g} = d\omega/dk$ of the guided mode vanishes at the edge $k=0$
(with $\omega a/2 \pi c = 0.3168$) of the propagation band. At small
wave vectors $(ka/2\pi) < 0.1$, it can be approximated as: $v_{g}/c
\approx 1.8155 (ka/2\pi) - 0.94776 (ka/2\pi)^2$. All numerical
results presented in this paper are obtained by employing the
Wannier functions approach~\cite{Busch:2003-R1233:JPCM} using eleven
maximally localized Wannier functions; they have also been checked
to be in an excellent agreement with the results based on the
plane-waves calculations~\cite{mpb}.

First, we emphasize that in general case light transmission at the
band edges of PhC waveguide structures is vanishing, due to
vanishing group velocity $v_{g}$. This effect is responsible, in
particular, for strong (scaling as $1/v_{g}$) extrinsic scattering
loss of slow light in PhC waveguides due to random fabrication
imperfections such as surface roughness and
disorder~\cite{Hughes:2005-033903:PRL}.

To illustrate, in Fig.~\ref{fig:fig2} we present the slow-light
transmission spectra for the waveguide-cavity structure based on the
W1-waveguide coupled to a cavity created by a single defect rod with
the radius $r_{\rm{def}}$. In what follows we refer to this
structure as a single-defect geometry. Changing $r_{\rm{def}}$, we
shift the resonance frequency $\omega_{\rm{res}}$ of this structure
from the middle of the propagation band, at $r_{\rm{def}}=0$, to the
edge ($k=0$) at $r_{\rm{def}} \approx 0.103a$. As was mentioned
above, for such a generic structure the light transmission vanishes
at the propagation band edges (at $\omega a/2 \pi c = 0.3168$ in our
case). This effect can be understood by analyzing an effective
discrete model of the waveguide-cavity structures
\cite{Mingaleev:2006-046603:PRE,Miroshnichenko:2005-036626:PRE}. In
such discrete model the transmission coefficient can be calculated
as $T(\omega)=\sigma^2(\omega)/[\sigma^2(\omega)+1]$, where the
function $\sigma(\omega)$ is determined by the structure geometry.
For a high-quality resonance with $\omega_{\rm{res}}$ lying deeply
inside the propagation band $\sigma(\omega) \simeq
\sigma_{\rm{Lorenz}}(\omega) \equiv 2Q(1-\omega/\omega_{\rm{res}})$
is determined by the resonance quality factor $Q$. However,
$\sigma(\omega)$ changes substantially near the band edges $k=0$ and
$k=\pi/s$, where $s$ is the waveguide period ($s=a$ for the W1
waveguide). For most of the structures, $\sigma(\omega)$ can be
approximated as $\sigma(\omega) \simeq \sin (ks)
\sigma_{\rm{Lorenz}}(\omega) \sim v_{g}
\sigma_{\rm{Lorenz}}(\omega)$ and, therefore, the transmission
coefficient {\em vanishes at the band edges}, where $v_{g} \to 0$.
This effect can be understood as an effective reduction of the
quality factor $Q \sim v_{g}$ in the slow-light regime, clearly seen
in Fig.~\ref{fig:fig2}. Moreover, when the resonance frequency
approaches the band edge, the maximally achievable (in the
slow-light region) transmission vanishes too (see
Fig.~\ref{fig:fig2}). Therefore, this structure cannot be employed
for the slow-light switching because the optical bistability in such
structures will become possible only
\cite{Mingaleev:2006-046603:PRE,Yanik:2003-2739:APL} when the linear
transmission exceeds 75\%.

\begin{figure}[t]
\centering\includegraphics[width=12cm]{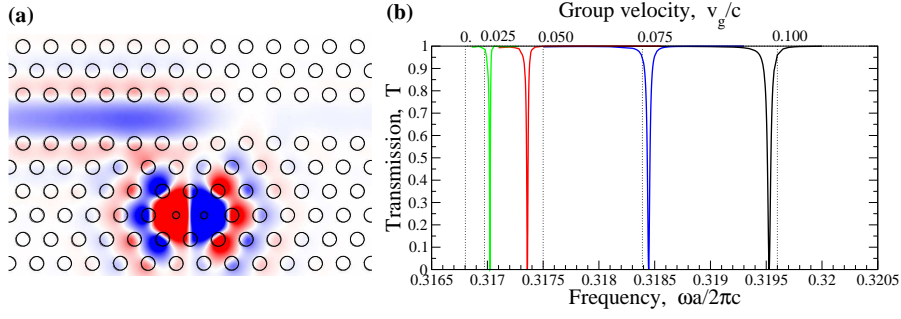} \caption{
Double-defect waveguide-cavity structure with the cavity created by
two defect rods with the radius $r_{\rm{def}}$: (a) Electric field
at the resonance reflection for $r_{\rm{def}}=0.121a$; (b)
Transmission spectra for different values of $r_{\rm{def}}$:
$0.119a$ (black), $0.120a$ (blue), $0.121a$ (red), $0.1213a$
(green).} \label{fig:fig3}
\end{figure}

\section{Quality factor enhancement and slow-light bistability}

We found that the effect of vanishing light transmission at the band
edges of PhC waveguides is very common, and it appears in a variety
of commonly used resonant structures. However, using the discrete
nature of PhCs, it becomes possible to design such waveguide-cavity
structures for which $\sigma(\omega) \simeq \tan (ks/2)
\sigma_{\rm{Lorenz}}(\omega)$ will grow inversely proportional with
the vanishing group velocity, $\sigma(\omega) \sim (1/v_{g})
\sigma_{\rm{Lorenz}}(\omega)$, at the band edge
$k=\pi/s$~\cite{Mingaleev:2006-046603:PRE}. At this band edge $T\to
1$ and the effective quality factor of the resonance should grow as
$Q \sim 1/v_{g}$ when the resonant frequency approaches the band
edge. This increase in $Q$ leads to significant lowering of the
bistability threshold for all-optical light switching in the
slow-light regime. Such a structure can be designed by placing a
side-coupled cavity between two nearest defects of a PhC waveguide
assuming that all the defect modes and the cavity mode have the same
symmetry. In Ref.~\cite{Mingaleev:2006-046603:PRE}, we illustrated
these results for the so-called coupled-resonator waveguides made by
removing every second rod; this example looks however a bit
artificial having limited applicability.

Our extensive analysis shows that the geometry engineering can be
employed effectively to achieve the slow-light switching  in
different types of PhC structures, including the important case of
the W1 waveguide and the propagation band edge at $k=0$. In a
general case, this approach is based on placing a side-coupled
cavity with an appropriate symmetry of the cavity mode into special
locations along the PhC waveguide. These locations and the mode
symmetry should be chosen in such a way that the overlap between the
cavity mode and guided mode at the band edge vanishes and,
consequently, scattering of light by the resonator at this band edge
vanishes too. As we already indicated, in the case when the
waveguide's defect modes have the same symmetry as the cavity mode,
such a vanishing of the modes' overlap is only possible at the band
edge $k=\pi/s$, leading to $\sigma(\omega) \simeq \tan (ks/2)
\sigma_{\rm{Lorenz}}(\omega)$. However, as can be shown by direct
extending of the results of the discrete model~
\cite{Miroshnichenko:2005-036626:PRE,Mingaleev:2006-046603:PRE}, in
the case of the opposite (even-odd) symmetry of the waveguide defect
modes and the side-coupled cavity mode, such a vanishing of the
modes' overlap becomes possible at the band edge $k=0$, leading to
$\sigma(\omega) \simeq \cot (ks/2) \sigma_{\rm{Lorenz}}(\omega)$.
Therefore, in this case both $\sigma(\omega)$ and the resonance
quality factor grow inversely proportional with vanishing group
velocity at the band edge $k=0$: $\sigma(\omega) \sim (1/v_{g})
\sigma_{\rm{Lorenz}}(\omega)$; and, therefore, we obtain $Q \sim
1/v_{g}$.

\begin{figure}[t]
\centering\includegraphics[width=11cm]{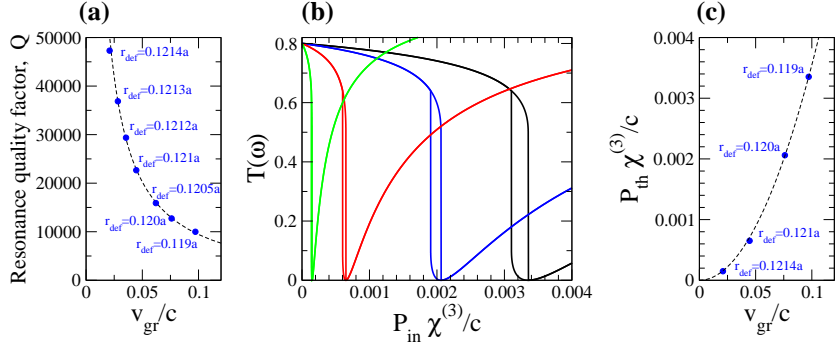} \caption{ (a)
Quality factor $Q$ vs. group velocity $v_{g}$ at resonance for the
structure shown in Fig.~\ref{fig:fig3}; (b) Nonlinear bistable
transmission in the same structure at the frequencies with 80\% of
linear light transmission vs. the incoming light power for different
values of $r_{\rm{def}}$: $0.119a$ (black), $0.120a$ (blue),
$0.121a$ (red), $0.1214a$ (green); (c) Switch-off bistability
threshold $P_{\rm{th}}$ vs. the group velocity $v_{g}$ at resonance
for the same structure.} \label{fig:fig4}
\end{figure}

This is exactly the case of the waveguide-cavity structure presented
in Fig.~\ref{fig:fig3} which utilizes odd-symmetry mode of a
double-defect cavity. Indeed, we use the even-symmetry defect modes
creating the waveguide and thus to achieve high-$Q$ resonance in the
slow-light regime at the band edge $k=0$ we should employ a
side-coupled cavity mode with the odd symmetry. The simplest cavity
of this type is created by reducing the radius $r_{\rm{def}}$ of
{\em two neighboring rods}, as shown in Figs.~\ref{fig:fig1}(b). The
range of the resonance frequencies of such a structure occupies
almost the whole propagation band from its upper boundary at
$r_{\rm{def}}=0$ to its edge $k=0$ at $r_{\rm{def}} \approx
0.1215a$. In Fig.~\ref{fig:fig3} we present the linear transmission
spectra for several values of the radius $r_{\rm{def}}$. As is seen,
in all the cases the light transmission remains perfect at the band
edge $k=0$ due to decoupling guided and cavity
modes~\cite{Mingaleev:2006-046603:PRE,Miroshnichenko:2005-036626:PRE}.
The resonance quality factor $Q$ grows when the resonance frequency
approaches the band edge. Numerically obtained dependence $Q(v_{gr})
\sim 1/v_{gr}$ is shown in Fig.~\ref{fig:fig4}(a), and it is in an
excellent agreement with the theoretical predictions. Since the
bistability threshold power of the incoming light in
waveguide-cavity structures scales as $P_{th} \sim 1/Q^2$
\cite{Mingaleev:2006-046603:PRE}, we should observe rapid vanishing
of $P_{th} \sim v_{g}^2$ when the resonance frequency approaches the
band edge. Indeed, direct numerical calculations summarized in
Figs.~\ref{fig:fig4}(b,c) prove this idea. These results were
calculated with the straightforward nonlinear extension of the
Wannier function approach \cite{Busch:2003-R1233:JPCM} using 11
maximally localized Wannier functions.

\section{Conclusions}

We have analyzed the resonant transmission and bistability of slow
light in a photonic-crystal waveguide coupled to a nonlinear cavity.
We have shown how to achieve the perfect transmission near the edges
of the propagation band by adjusting either the cavity location
relative to the waveguide or the mode symmetry. We emphasize that
such properly designed photonic structures are suitable for the
observation of high-$Q$ resonant (with the quality factor $Q\sim
1/v_{\rm g}$) and ultralow-threshold bistable (with threshold power
$P_{\rm th} \sim v_{\rm g}^2$) transmission of slow light with the
small group velocity $v_{\rm g}$ at the resonance frequency. Thus,
engineering the geometry and mode symmetry of the photonic-crystal
structures is an useful tool for developing novel concepts of
all-optical switching devices operating in the slow-light regime.

\vspace{5mm}

{\bf Acknowledgments.} The work was supported by the Australian
Research Council and the National Academy of Sciences of Ukraine
through the Fundamental Research Program.


\begin{thebibliography}{99}

\bibitem{Dowling:1994-1896:JAP}
J.P. Dowling, M. Scalora, M.J. Bloemer, and C.M. Bowden, ``The
photonic band edge laser: A new aproach to gain enhancement,'' J.
Appl. Phys. {\bf 75,} 1896--1899 (1994).

\bibitem{Roussey:2006-241110:APL}
M. Roussey, M.-P. Bernal, N. Courjal, D. Van Labeke, F.I. Baida, and
R. Salut, ``Electro-optic effect exaltation on lithium niobate
photonic crystals due to slow photons,'' Appl. Phys. Lett. {\bf 89,}
241110 (2006).

\bibitem{Suzuki:1995-570:JOSB}
T. Suzuki and P.K.L. Yu, ``Emission power of an electric dipole in
the photonic band structure of the fcc lattice,'' J. Opt. Soc. Am. B
{\bf 12,} 570--582 (1995).

\bibitem{Scalora:1994-1368:PRL}
M. Scalora, J.P. Dowling, C.W. Bowden, and M.J. Bloemer, ``Optical
limiting and switching of ultrashort pulses in nonlinear photonic
band gap materials,'' Phys. Rev. Lett. {\bf 73,} 1368--1371 (1994).

\bibitem{Martorell:1997-702:APL}
J. Martorell, R. Vilaseca, and R. Corbalan, ``Second-harmonic
generation in a photonic crystal,'' Appl. Phys. Lett. {\bf 70,}
702--704 (1997).

\bibitem{Soljacic:2002-2052:JOSB}
M. Soljacic, S.G. Johnson, S.H. Fan, M. Ibanescu, E. Ippen, and J.D.
Joannopoulos, ``Photonic-crystal slow-light enhancement of nonlinear
phase sensitivity,'' J. Opt. Soc. Am. B {\bf 19,} 2052--2059 (2002).

\bibitem{Chen:2004-3353:OE}
Y. Chen and S. Blair, ``Nonlinearity enhancement in finite
coupled-resonator slow-light waveguides,'' Opt. Express {\bf 12,}
3353--3366 (2004).

\bibitem{Khurgin:2005-1062:JOSB}
J.B. Khurgin, ``Optical buffers based on slow light in
electromagnetically induced transparent media and coupled resonator
structures: comparative analysis,'' J. Opt. Soc. Am. B {\bf 22,}
1062--1074 (2005).

\bibitem{Xia:2007-65:NATPHOT}
F. Xia, L. Sekaric, and Yu. Vlasov, ``Ultracompact optical buffers
on a silicon chip,'' Nat. Photon. {\bf 1,} 65--72 (2007).

\bibitem{Notomi:2001-253902:PRL}
M. Notomi, K. Yamada, A. Shinya, J. Takahashi, C. Takahashi, and I.
Yokohama, ``Extremely large group velocity dispersion of line-defect
waveguides in photonic crystal slabs,'' Phys. Rev. Lett. {\bf 87,}
253902 (2001).

\bibitem{Jacobsen:2005-7861:OE}
R.S. Jacobsen, A.V. Lavrinenko, L.H. Frandsen, C. Peucheret, B.
Zsigri, G. Moulin, J.F. Pedersen, and P. I. Borel, ``Direct
experimental and numerical determination of extremely high group
indices in photonic crystal waveguides,'' Opt. Express {\bf 13,}
7861--7871 (2005).

\bibitem{Vlasov:2005-65:NAT}
Y.A. Vlasov, M. O'Boyle, H.F. Hamann, and S.J. McNab, ``Active
control of slow light on a chip with photonic crystal waveguides,''
Nature {\bf 438,} 65--69 (2005).

\bibitem{Gersen:2005-073903:PRL}
H. Gersen, T.J. Karle, R.J.P. Engelen, W. Bogaerts, J.P. Korterik,
N.F. van Hulst, T.F. Krauss, and L. Kuipers, ``Near-field
characterization of low-loss photonic crystal waveguides,'' Phys.
Rev. Lett. {\bf 94,} 073903 (2005).

\bibitem{Assefa:2006-745:OL}
S. Assefa, S.J. McNab and Y.A. Vlasov, ``Transmission of slow light
through photonic crystal waveguide bends,'' Opt. Lett. {\bf 31,}
745--747 (2006).

\bibitem{Vlasov:2006-50:OL}
Y.A. Vlasov and S.J. McNab, ``Coupling into the slow light mode in
slab-type photonic crystal waveguides,'' Opt. Lett. {\bf 31,} 50--52
(2006).

\bibitem{Mok:2006-775:NATPHYS}
J.T. Mok, C.M. de Sterke, I.C.M. Littler, and B.J. Eggleton,
``Dispersionless slow light using gap solitons,'' Nat. Phys. {\bf
2,} 775--780 (2006).

\bibitem{Fan:2002-908:APL}
S.H. Fan, ``Sharp asymmetric line shapes in side-coupled
waveguide-cavity systems,'' Appl. Phys. Lett. {\bf 80,} 908--910
(2002).

\bibitem{Almeida:2004-1081:NAT}
V. R. Almeida, C. A. Barrios, R. R. Panepucci, and M. Lipson,
``All-optical control of light on a silicon chip,'' Nature {\bf
431,} 1081--1084 (2004).

\bibitem{Barclay:2005-801:OE}
P. E. Barclay,  K. Srinivasan, and O. Painter, ``Nonlinear response
of silicon photonic crystal microresonators excited via an
integrated waveguide and fiber taper,'' Opt. Express {\bf 13,}
801--820 (2005).

\bibitem{Notomi:2005-2678:OE}
M. Notomi, A. Shinya, S. Mitsugi, G. Kira, E. Kuramochi, and T.
Tanabe, ``Optical bistable switching action of Si high-Q
photonic-crystal nanocavities,'' Opt. Express {\bf 13,} 2678--2687
(2005).

\bibitem{Tanabe:2005-151112:APL}
T. Tanabe, M. Notomi, S. Mitsugi, A. Shinya, and E. Kuramochi,
``All-optical switches on a silicon chip realized using photonic
crystal nanocavities,'' Appl. Phys. Lett. {\bf 87,} 151112 (2005).

\bibitem{Priem:2005-9623:OE}
G. Priem, P. Dumon, W. Bogaerts, D. Van Thourhout, G. Morthier, and
R. Baets, ``Optical bistability and pulsating behaviour in
Silicon-On-Insulator ring resonator structures,'' Opt. Express {\bf
13,} 9623--9528 (2005).

\bibitem{Uesugi:2006-377:OE}
T. Uesugi, B. Song, T. Asano, and S. Noda, ``Investigation of
optical nonlinearities in an ultra-high-Q Si nanocavity in a
two-dimensional photonic crystal slab,'' Opt. Express {\bf 14,}
377--386 (2006).

\bibitem{Yang:2007-0703132:arXiv}
X. Yang, C. Husko, M. Yu, D.-L. Kwong, and C.W. Wong, ``Observation
of femto-joule optical bistability involving Fano resonances in
high-Q/$V_{m}$ silicon photonic crystal nanocavities,''
arXiv:physics/0703132 (2007).

\bibitem{Hughes:2005-033903:PRL}
S. Hughes, L. Ramunno, J.F. Young, and J.E. Sipe, ``Extrinsic
optical scattering loss in photonic crystal waveguides: Role of
fabrication disorder and photon group velocity,'' Phys. Rev. Lett.
{\bf 94,} 033903 (2005).

\bibitem{Mingaleev:2006-046603:PRE}
S.F. Mingaleev, A.E. Miroshnichenko, Y.S. Kivshar, and K. Busch,
``All-optical switching, bistability, and slow-light transmission in
photonic crystal waveguide-resonator structures,'' Phys. Rev. E {\bf
74,} 046603 (2006).

\bibitem{Busch:2003-R1233:JPCM}
K. Busch, S.F. Mingaleev, A. Garcia-Martin, M. Schillinger, and D.
Hermann, ``Wannier function approach to photonic crystal circuits,''
J. Phys.: Condens. Matter. {\bf 15,} R1233--R1256 (2003).

\bibitem{mpb}
%S.G. Johnson, ``MIT Photonic-Bands package,''
%http://ab-initio.mit.edu/mpb/
S.G. Johnson and J.D. Joannopoulos, ``Block-iterative
frequency-domain methods for Maxwell's equations in a planewave
basis,'' Opt. Express {\bf 8,} 173--190 (2001).

\bibitem{Miroshnichenko:2005-036626:PRE}
A.E. Miroshnichenko, S.F. Mingaleev, S. Flach, and Yu.S. Kivshar,
``Nonlinear Fano resonance and bistable wave transmission,'' Phys.
Rev. E {\bf 71,} 036626 (2005).

\bibitem{Yanik:2003-2739:APL}
M.F. Yanik, S.H. Fan, and M. Soljacic, ``High-contrast all-optical
bistable switching in photonic crystal microcavities,'' Appl. Phys.
Lett. {\bf 83,} 2739--2741 (2003).

\end{thebibliography}
\end{document}